\documentclass{article}
\usepackage{spconf,amsmath,graphicx}

\usepackage{amssymb}
\usepackage{amsmath}
\usepackage{tikz}
\usetikzlibrary{graphs}
\usepackage{subfigure}
\usepackage{verbatim}
\usepackage{amsfonts}

\usepackage{float}

\title{Attention-based Transfer Learning for Brain-computer Interface}
\name{Chuanqi Tan \qquad Fuchun Sun \qquad Tao Kong \qquad Bin Fang \qquad Wenchang Zhang
\thanks{This work is jointly supported by National Natural Science Foundation of China under with Grant No.91848206, 61621136008, U1613212.}}
\address{State Key Laboratory of Intelligent Technology and Systems
    \\Tsinghua National Laboratory for Information Science and Technology (TNList)
    \\Department of Computer Science and Technology, Tsinghua University\\
    \{tcq15@mails, fcsun@mail, kt14@mails, fangbin@mail, zhangwc14@mails\}.tsinghua.edu.cn}

\begin{document}
\maketitle
\begin{abstract}
    
Different functional areas of the human brain play different roles in brain activity, which has not been paid sufficient research attention in the brain-computer interface (BCI) field.
This paper presents a new approach for electroencephalography (EEG) classification that applies attention-based transfer learning. 
Our approach considers the importance of different brain functional areas to improve the accuracy of EEG classification, and provides an additional way to automatically identify brain functional areas associated with new activities without the involvement of a medical professional. 
We demonstrate empirically that our approach out-performs state-of-the-art approaches in the task of EEG classification, and the results of visualization indicate that our approach can detect brain functional areas related to a certain task.

\end{abstract}

\section{Introduction}

Brain-computer interface (BCI) based systems can read the brain information of the subject and decode it into instructions for controlling an external device, thereby interacting more naturally with the user. 
The key issue in BCI-based systems is the accuracy of Electroencephalography (EEG) classification.

One of the most important problems in EEG classification methods is that \textit{the relationship between the functional areas of the human brain and the specific activities is not effectively utilized}. 
This problem makes it difficult to find key electrodes with higher signal-to-noise ratios, therefore, it is difficult to obtain an effective EEG classifier. 
Medical research has shown that the functional areas of the human brain have strong regional correlation with specific activities. 
In the past, when we designed BCI-based systems, we needed medical experts to specify key electrodes involved in special activity. 
For a new activity, it is difficult to construct a usability system without the help of medical experts, which severely limits the applicability of  BCI-based systems. 
It would be meaningful to have a non-medical approach that automatically discovers activity-related functional areas from brain signals.
In addition, another key issue is the lack of training data. Because the cost of biosignal acquisition and labeling is extremely high, it is almost impossible to construct a large, high-quality EEG signal dataset. It is difficult to train advanced classifiers without sufficient training samples.

\begin{figure}[t]
    \centering
    \includegraphics[page=1,trim=.4cm 9.4cm 21.3cm 9.5cm,clip,width=\columnwidth]{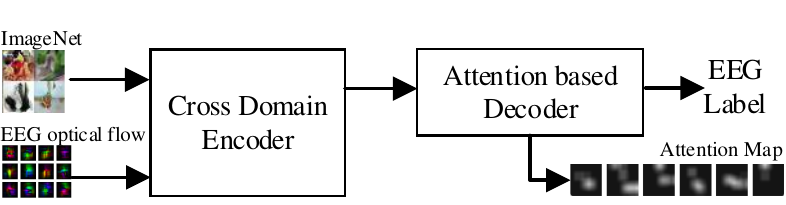}
    
    \caption{Overview of our approach. 
        We applied ImageNet as the source domain and EEG optical flow as the target domain to an attention-based transfer learning framework.
        In addition to obtain EEG label, it gets an extra \textbf{attention map} to reflect the activity of the human brain.}
    \label{pipeline}
\end{figure}

To solve these problems, we propose an \textit{attention-based transfer learning framework} that includes two main components: a cross domain encoder and an attention-based decoder with recurrent neural network (RNN). An overview of our approach is shown in Figure \ref{pipeline}.
A cross domain encoder has the ability to transfer knowledge from natural images domain by representing the original EEG signal in a new form - EEG optical flow. 
It uses a large amount of training data in the source domain (image classification) to help train the complex feature extractor in the target domain (EEG classification), which solves the problem of a lack of training data by using adversarial transfer learning. 
The feature extractor will be transferred as knowledge to the target domain.
An attention-based decoder uses the attention mechanism to automatically discover the weights of the brain functional areas, which effectively improves the accuracy of EEG classification. This mechanism can reflect the brain functional areas related to a specific activity and overcomes the reliance on medical experts when dealing with new activity.

The main \textbf{contributions} of this paper are as follows:
(1) We introduce attention-based transfer learning to the EEG classification task.  
(2) Our approach provides a novel way to automatically discover brain functional areas associated with new activities, reducing reliance on medical experts.
(3) Experiments show that our approach out-performs the state-of-the-art approaches in an EEG classification task and verify the usability of our approach.

\section{Related Work}

Many works have been conducted to improve EEG classification accuracy 
and a great variety of hand-designed features have been proposed.
With the rapid development of deep learning in recent years,  many excellent networks have been presented by researchers.
In recent years, many public works have discussed deep learning applications in bioinformatics research \cite{Mamoshina2016Applications}.

Transfer learning \cite{pan2010survey} and deep transfer learning \cite{tan2018survey} enable the use of different domains, tasks, and distributions for training and testing. 
\cite{jayaram2016transfer} reviewed the current state-of-the-art transfer learning approaches in BCI.
\cite{tan2017multimodal} proposed a novel EEG representation that reduces the EEG classification problem to an image classification problem that implicates the ability of transfer learning.
\cite{hajinoroozi2017deep} transferred general features via a convolutional network across subjects and experiments. 
\cite{lin2017improving} evaluated the transferability between subjects by calculating distance and transferred knowledge in comparable feature spaces to improve accuracy.
\cite{tan2018deep} designed a deep transfer learning framework which is suitable for transferring knowledge by joint training.

\cite{rensink2000dynamic} and \cite{corbetta2002control} discussed whether the human visual system has attention mechanism. 
\cite{wang2016survey} reviewed the recent works on attention-based RNN and its application in computer vision, and categorized the approaches into four classes: item-wise soft attention, item-wise hard attention, location-wise hard attention, and location-wise soft attention.
\cite{mnih2014recurrent} applied the visual attention mechanism in an RNN network to obtain the ability to extract information from images or video by adaptively selecting a sequence of regions or locations.
In \cite{ba2014multiple}, an attention-based model is applied to identify multiple objects in an image by using reinforcement learning to identify the most relevant regions of the input image. 
\cite{mnih2014recurrent} demonstrated that attention not only works on object detection tasks but many other computer vision tasks like image classification.
\cite{kiros2014multimodal} introduced an attention-based model to an image caption task.
\cite{xu2015show} proposed to extract the feature vector by using the intermediate layer of VGG, and the feature can be associated with a specific region in the image through the network map.
In natural language processing tasks, \cite{bahdanau2014neural} applied a soft attention mechanism to machine translation. \cite{vaswani2017attention} showed the latest attention model Google use in machine translation, which uses only attention without a convolutional neural network (CNN) or an RNN in a traditional encoder-decoder model.

To the best of our knowledge, no researchers have attempted to automatically discover brain functional areas associated with new activities.

\section{Method}

Our approach has a traditional encoder-decoder structure that consists of \textbf{two main components}: a cross domain encoder and an attention-based decoder.

\subsection{Cross domain encoder}

To obtain the ability of transfer learning, the raw EEG signal was converted to a new representation - EEG optical flow, which was proposed in our previous work \cite{tan2017multimodal}. 
Many benefits can be gained from using the EEG optical flow. 
In particular, EEG optical flow can \textit{enhance the ability of transfer learning from natural images}.

Many studies have demonstrated that the front layers in a convolutional neural network (CNN) can extract the general features of images, such as edges and corners.
Therefore, we were able to transfer the front layers of a CNN network trained on ImageNet to extract the general features of the EEG optical flow.
However, the general feature extractor trained by natural images does not fully match the EEG optical flow.

\begin{figure}[h]
    \centering
    \includegraphics[page=1,trim=1.cm 18.8cm 20.5cm .5cm,clip,width=\columnwidth]{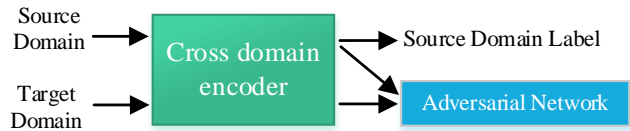}
    
    \caption{Pipeline of adversarial transfer learning.Adversarial network used to identify the origins of input features.}
    \label{pipeline of adversarial transfer learning}
\end{figure}

Inspired by generative adversarial nets (GAN), we apply an adversarial network to train a better general feature extractor which is described in our previous work \cite{tan2018deep}. The pipeline of adversarial transfer learning is shown in Figure \ref{pipeline of adversarial transfer learning}.

We use features extracted from natural images and the EEG optical flow as the inputs for the adversarial network and train it to identify their origins. If the adversarial network achieves inferior performance, it indicates a small difference between the two types of feature and better transferability, and vice versa.
It can be achieved by optimizing this loss function:
\begin{equation}
\label{eq:loss3} \mathcal{L} = -\sum_{k}{ \mathbb{I} [y=k]\log p_{k}} + \alpha \mathcal L_{adver} + \beta \Re(v), \\
\end{equation}
where $k$ is the number of categories, $p_{k}$ is the softmax value of the classifier activations, 
$\mathcal L_{adver}$ is the cross entropy of the adversarial network, $\Re(v)$ is the regularization of manifold constraints, and $\alpha$ and $\beta$ are hyperparameters. 

Manifold constraints force the learning algorithm to transfer useful knowledge from the source domain and ignore the knowledge which may destroy the manifold structure of the target domain.
\cite{cai2011graph} demonstrated that keeping the geometric structure can be reduced to the regularization of:

\begin{equation}
\Re(v) = \dfrac{1}{2} \sum_{i,j=1}^{n} \zeta(v_{i*}, v_{j*})(W)_{ij}
\end{equation}
where $v_i$ is the embedded representation of sample $x_i$, $\zeta(v_{i*}, v_{j*})$ is the loss function to measure the euclidean distance of $v_{i*}$ and $v_{j*}$, $(W)_{ij}$ is the cosine similarity measure of $p-nearest$ neighbor in the adjacency matrix.

 To train this adversarial network, we applied an \textit{iteratively optimizing algorithm} with two steps, which has been described in our previous work \cite{tan2018deep}.
In this section, we have introduced a cross domain encoder that extracts features suitable for both the source and target domains and obtains the high-quality features of an EEG signal with help from natural images.

\subsection{Attention-based decoder}

In this section, we use the features extracted by the cross domain encoder to obtain the final EEG label and attention map of the brain through the attention-based decoder. 
The attention-based decoder is an RNN network, and we feed the features obtained from each EEG optical flow frame into the RNN network, and treat the output of the last timestamp as the final EEG label.

In a traditional encoder-decoder network, the input of the decoder is the output of the last fully connected layer of the encoder, which raises a crucial problem. 
The features extracted from the last fully connected layer \textit{lose the location information} of the brain functional areas, so these features do not reflect the importance of different brain functional areas for specific activities.

\begin{figure*}[t]
    \centering
    \includegraphics[page=1,trim=0.6cm 14.cm 12.8cm 3.6cm,clip,width=0.9\textwidth]{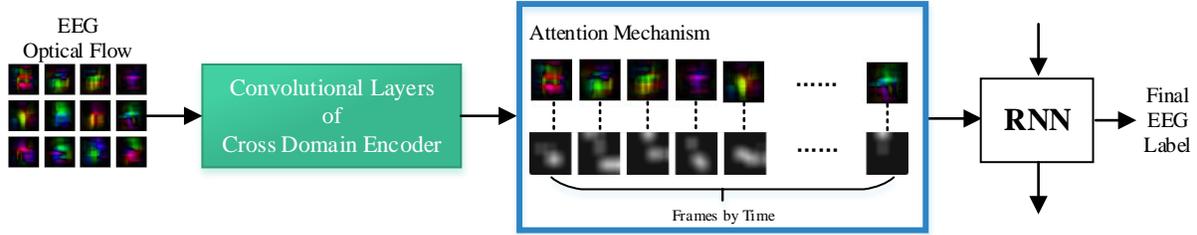}
    
    \caption{Pipeline of attention-based decoder. 
        First, the feature vectors that can maintain spatial information are produced by the convolutional layers of the cross domain encoder. 
        Then, they are combined with the location-wise attention mechanism on each frame. Finally, these feature vectors are sent to the RNN network one by one to obtain the final EEG label.}
    \label{pipeline of decoder}
\end{figure*}

Encouraged by recent works in computer vision and neural language process, and inspired by recent success in employing attention in these research works, we applied an attention-based decoder that can attend to salient parts of an EEG optical flow while carrying out EEG classification.
The attention mechanism provided a powerful tool to overcome the important issue mentioned above.
The \textit{location-wise attention mechanism} allows us to consider the weight of different parts in the EEG optical flow, which reflect the different functional areas of the human brain.
The pipeline of our attention-based decoder is shown in Figure \ref{pipeline of decoder}.

The encoder can obtain feature vectors of each EEG optical flow. In order to link the items in the feature vector to the parts of the EEG optical flow one by one, we use the feature map of the \textit{convolutional layer} instead of the output of the \textit{fully connected layer}. Since a low-level feature retains more information, it will be lost in the fully-connected layer. In this way, we can extract $L$ vector features of $D$ dimension as feature vectors, each dimension of feature vector corresponding to a part of the EEG optical flow, as shown in the following equation:
\begin{equation}
a = \{\text{\textbf{a}}_1, ... , \text{\textbf{a}}_L\}, \text{\textbf{a}}_i \in \mathbb{R}^D ,
\end{equation}
where $L$ is the number of frames and $D$ is the number of areas on the EEG optical flow.
The items of feature vector are linked to the spatial location of the EEG optical flow by convolutional operation, which is demonstrated in Figure \ref{corresponds of spatial location}:

\begin{figure}[H]
    \centering
    \begin{tikzpicture}
    \node[anchor=south west,inner sep=0, label=below:\footnotesize EEG optical flow] (a) at (0.,0.) {\includegraphics[width=.45in]{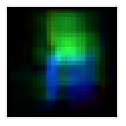}};  
    
    \node (b) at ( 3.7,0.55) [rectangle] {$\{\text{\textbf{a}}_1, \text{\textbf{a}}_2, ... , \text{\textbf{a}}_L\}$};
    
    \draw[yellow,line width=.5mm] (.2,0.95) .. controls (1., 1.7) .. (2.85,0.65);
    \draw[green,line width=.5mm] (.5,0.95) .. controls (1., 1.5) and (2.5, 1.5) .. (3.4,0.66);
    \draw[red,line width=.5mm] (.95,0.2) .. controls (2., -0.2) and (4., -0.2) .. (4.4,0.4);
    \end{tikzpicture}
    
    \caption{Link between items in the feature vector and spatial location of the EEG optical flow.}
    \label{corresponds of spatial location}
\end{figure}
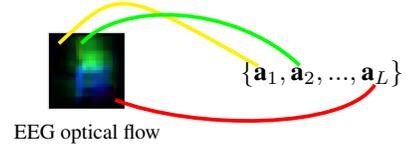

In the attention mechanism, we need to obtain the context vector as the input of the RNN at each time $t$. The following equations are applied to calculate the context vector: $\text{\^{z}} _t$:
\begin{eqnarray}
e_{ti} =& f_{att}(\text{\textbf{a}}_i, h_{t-1}) \\
\alpha_{ti} =& \dfrac{exp(e_{ti})}{\sum_{k=1}^{L} exp(e_{tk})} \\
\text{\^{z}} _t =& \phi (\{\textbf{a}_i\}, \{\alpha_i\}),
\end{eqnarray}
where $h_{t-1}$ is the hidden state of the previous step, $f_{att}$ is the map of a multilayer perceptron (MLP), $e_{ti}$ is the output of the MLP, $\alpha_{ti}$ is the attention weights and $\phi$ is the function combining feature vectors and attention weights.

There are two types of attention mechanism, the soft attention mechanism and the hard attention mechanism. 
The main difference is the definition of the $\phi$ function. In the soft attention mechanism, $\phi (\{\textbf{a}_i\}, \{\alpha_i\}) = \sum_{i}^{L} \alpha_i \textbf{a}_i$ that means all parts of the EEG optical flow will be considered in the context vector $\text{\^{z}} _t$. In the hard attention mechanism, $\phi$ is a function that returns a sampled $\textbf{a}_i$ at every point in time according to the multinouilli distribution parameterized by $\alpha$ .

The soft attention mechanism is a smooth function; it can be solved directly by the back propagation algorithm, which is equivalent to optimizing the following loss function:
\begin{equation}
\mathcal{L} = -log(P(y|x)) + \alpha \sum_{i}^{L} (1- \sum_{t}^{C} \alpha_{ti})^2.
\end{equation}
The hard attention mechanism is a non-smooth function that can be approximated by the Monte Carlo algorithm.

\section{Experiments}

We applied our approach to a dataset called Open Music Imagery Information Retrieval (OpenMIIR) \cite{stober2017learning}.
OpenMIIR is compiled during music perception and imagination, which involves 10 subjects listening to and imagining 12 short music fragments taken from well-known pieces. 
These signals were recorded using 64 EEG electrodes at 512 Hz, and 240 trials were recorded per subject.
The following parameters were used in our approach. We converted raw EEG signals into EEG videos with thirteen frames and a resolution of 32*32. These frames were converted to EEG optical flow with twelve frames.
We employed VGG16 and VGG19 \cite{guo2016deep} as the targets of the cross domain encoder .

The OpenMIIR dataset does not distinguish between training and test sets, so we randomly selected 10\% of the dataset to use as the test dataset.
As the baseline, we tested some recently proposed approaches: the deep neural network (DNN) described in \cite{stober2017learning} and the CNN described in \cite{stober2015deep}.
In addition, we made comparisons to our previous work \cite{tan2018deep}, that without an attention mechanism.
Experiments on the OpenMIIR dataset were conducted to compare the performance of our approach and that of the baseline approaches, and the results are shown in Table \ref{table: classification accury on OpenMIIR}.

\begin{table}[h]
    \centering
    \caption{Classification accuracy (\%) on the OpenMIIR dataset and comparisons to the baseline approaches. For example, the corner mark in $Our_{(Soft+VGG16)}$ refers to use of the soft attention mechanism and application of the VGG16 network as the encoder.}
    \label{table: classification accury on OpenMIIR}
    \begin{tabular}{ll|ll}
        \hline
        \noalign{\smallskip}
        \cite{stober2017learning} & 27.22 & \cite{stober2015deep} & 27.80 \\ 
        \hline
        $\cite{tan2018deep}_{VGG16}$ & 32.08 & $\cite{tan2018deep}_{VGG19}$ & 35.00 \\ 
        \hline
        $Our_{(Soft+VGG16)}$ & 37.92 & $Our_{(Soft+VGG19)}$ & 36.67 \\ 
        \hline
        $Our_{(Hard+VGG16)}$ & 37.08 & $Our_{(Hard+VGG19)}$ & 35.84 \\ \noalign{\smallskip}
        \hline
    \end{tabular}
\end{table}

\begin{figure}[h]
    \centering
    \begin{minipage}[h]{1.2in} 
        \centering 
        
        \subfigure[$Soft$]{
            \includegraphics[width=0.39in]{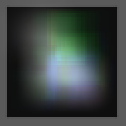}
        }
    \end{minipage}%
    \begin{minipage}[h]{1.2in} 
        \centering 
        
        \subfigure[$Hard$]{
            \includegraphics[width=0.39in]{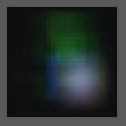}
        }
    \end{minipage}%
    
    \caption{Visualization of soft attention mechanism and hard attention mechanism on the same frame.}
    \label{different between soft and hard attention}
\end{figure}

As the results show in Table \ref{table: classification accury on OpenMIIR}, the soft attention mechanism achieves better classification results than the hard attention mechanism. One possible reason is that the soft attention mechanism considers the interaction between multiple functional areas, while the hard mechanism only considers one functional area, as shown in Figure \ref{different between soft and hard attention}. Medical knowledge tells us that the reflection of an activity in the brain is the result of a combination of multiple functional areas, which is more in line with the soft attention mechanism.

\begin{figure}[h]
    \centering
    \includegraphics[width=\columnwidth]{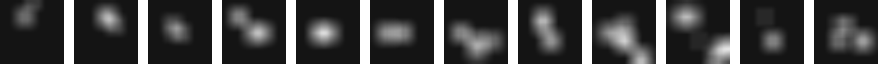}
    
    \caption{Visualization of soft attention mechanism when listing to an intense music fragment.}
    \label{visualization of 12 frames attention images}
\end{figure}

We visualized an attention map while a subject was listening to an intense piece of music, as shown in Figure \ref{visualization of 12 frames attention images}. It was found that the learned weights of attention are somewhat similar to the result from medical experts \cite{towle1993spatial}.

We can draw the following conclusions from the experimental results presented in this section:
(1) The experimental results shown in Table \ref{table: classification accury on OpenMIIR} demonstrate that our proposed approach performs better than traditional approaches;
(2) VGG16 is a better choice for encoder than VGG19 in our attention-based transfer learning for EEG classification task;
(3) The performance of the soft attention mechanism is better than that of the hard attention mechanism;
(4) Attention mechanisms can be used to automatically discover brain functional areas associated with new activities and reduce the dependence on medical experts.

\section{Conclusions}

We propose a novel approach to improve the accuracy of EEG classification in BCI. This approach takes advantage of the medical fact that different brain functional areas play different roles in activities.
It applies an attention mechanism to automatically assess the importance of functional areas of the brain during activity. 
It can be concluded that our approach is superior to other state-of-the-art approaches.
In addition, our approach can be used to automatically discover brain functional areas associated with activities, which is very useful when dealing with EEG data related to a new activity.

\bibliographystyle{IEEEbib}
\bibliography{icassp2019_atten_reference}

\end{document}